\documentclass{article}

    \PassOptionsToPackage{numbers, compress}{natbib}

\usepackage[final]{neurips_2024}

\usepackage[utf8]{inputenc} 
\usepackage[T1]{fontenc}    
\usepackage{hyperref}       
\usepackage{url}            
\usepackage{booktabs}       
\usepackage{amsfonts}       
\usepackage{nicefrac}       
\usepackage{microtype}      
\usepackage{xcolor}         
\usepackage{graphicx}
\usepackage{amsmath}
\usepackage{caption}
\usepackage{subcaption}

\title{SNAC: Multi-Scale Neural Audio Codec}

\author{%
  Hubert Siuzdak \\
  Papla Media \\
  \texttt{hubert@papla.media} \\
  \And
  Florian Gr\"otschla \\
  ETH Zurich \\
  \texttt{fgroetschla@ethz.ch} \\
  \AND
  Luca A. Lanzend\"orfer \\
  ETH Zurich \\
  \texttt{lanzendoerfer@ethz.ch} \\
}

\begin{document}

\maketitle

\begin{abstract}
Neural audio codecs have recently gained popularity because they can represent audio signals with high fidelity at very low bitrates, making it feasible to use language modeling approaches for audio generation and understanding. Residual Vector Quantization (RVQ) has become the standard technique for neural audio compression using a cascade of VQ codebooks. 
This paper proposes the Multi-Scale Neural Audio Codec, a simple extension of RVQ where the quantizers can operate at different temporal resolutions. By applying a hierarchy of quantizers at variable frame rates, the codec adapts to the audio structure across multiple timescales. This leads to more efficient compression, as demonstrated by extensive objective and subjective evaluations. The code and model weights are open-sourced at \url{https://github.com/hubertsiuzdak/snac}.
\end{abstract}

\section{Introduction}

Neural audio compression methods have recently achieved lower bitrates than traditional codecs while maintaining competitive quality \cite{zeghidour2021soundstream, kumar2024high}. By representing audio using discrete latent variables, these methods have proven useful beyond compression, particularly in generative models. This approach has led to successful generation of natural and coherent continuations of speech and music~\cite{borsos2023audiolm}.

However, existing audio tokenizers, while capable of high reconstruction quality, suffer from high token granularity that limits their ability to capture long-term structures. Specifically, in transformer architectures \cite{vaswani2017attention}, processing extended attention context windows becomes impractical. Ideally, discrete audio tokens should also represent higher-level aspects of sounds \cite{zhang2024speechtokenizer}. \emph{Semantic} tokens are commonly used in audio generation systems due to their higher information density, which enables autoregressive models to recognize patterns more reliably, much like in natural language processing. In contrast, high-frequency acoustic tokens often contain noise and less patterned data.

Audio signals inherently involve multiple levels of abstraction. Speech can be analyzed at local acoustic or phonetic levels, as well as in terms of prosody or semantics. Music also exhibits hierarchical structures over longer time scales. Research on the human auditory cortex also indicates that acoustic signals are processed hierarchically \cite{balaguer2009understanding, peelle2010hierarchical}.

In this paper, we introduce \textbf{SNAC} (Multi-\textbf{S}cale \textbf{N}eural \textbf{A}udio \textbf{C}odec), a simple extension to the current audio residual quantization approach by introducing quantization at different temporal resolutions to form a multi-scale discrete representation of audio. Our experiments -- including both objective metrics and subjective evaluations -- demonstrate that the proposed method achieves more efficient compression. Additionally, we introduce minor enhancements to the RVQGAN framework by incorporating noise blocks, depthwise convolutions, and local windowed attention. 

\section{Related Work}


\paragraph{Neural Audio Codecs}

Recent advancements in audio codecs have leveraged deep learning to learn efficient audio representations, moving beyond traditional signal processing methods \cite{kankanahalli2018end}. Vector quantization (VQ) \cite{gray1984vector}, which maps high-dimensional data onto a discrete set of code vectors, has been a foundational tool in compression. It was later applied in deep learning, leading to powerful discrete latent representations in VQ-VAE model \cite{van2017neural}. Successful applications of VQ-VAE in audio coding have been demonstrated \cite{garbacea2019low}. However, VQ becomes inefficient at higher bitrates due to the rapid growth of the codebook size. This issue was addressed early on by structuring the quantizer into multiple stages \cite{juang1982multiple}. This approach has now been effectively applied in modern end-to-end models, introduced as Residual Vector Quantization (RVQ) \cite{zeghidour2021soundstream}, enabling high-quality audio compression with scalable bitrates. Further advancements in RVQ have improved both the efficiency and quality of neural audio compression \cite{defossez2022high, kumar2024high}.

\paragraph{Multi-Scale Models}

Capturing long-term structure in generative models has proven challenging, and various approaches have been proposed to address this issue. A hierarchical decoder was introduced to manage the diffuculties of modeling long-term musical structures \cite{roberts2018hierarchical}. Subsequently, multi-level models for discrete sequences were explored \cite{dieleman2018challenge}. The multi-scale hierarchical organization of vector quantized codes was proposed to model large images effectively \cite{razavi2019generating}. Additionally, separate VQ-VAE models with different temporal resolutions were used to represent longer dependencies in musical compositions \cite{dhariwal2020jukebox}.

\section{Method}

\begin{figure}[t]
    \centering
    \begin{subfigure}{0.49\textwidth}
        \centering
        \includegraphics[height=7cm]{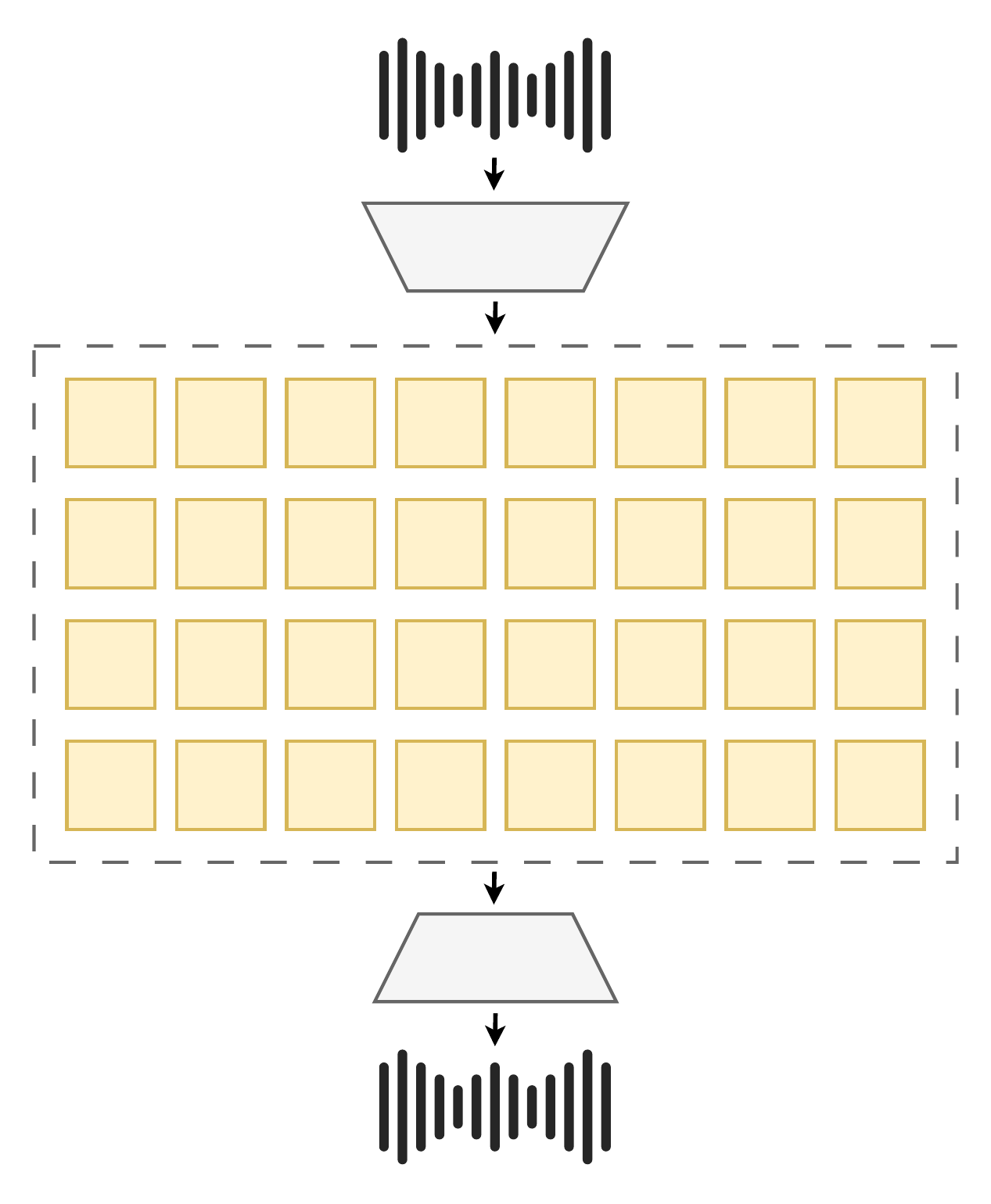}
        \caption{Residual Vector Quantization}
        \label{fig:rvq1}
    \end{subfigure}
    \hfill
    \begin{subfigure}{0.49\textwidth}
        \centering
        \includegraphics[height=7cm]{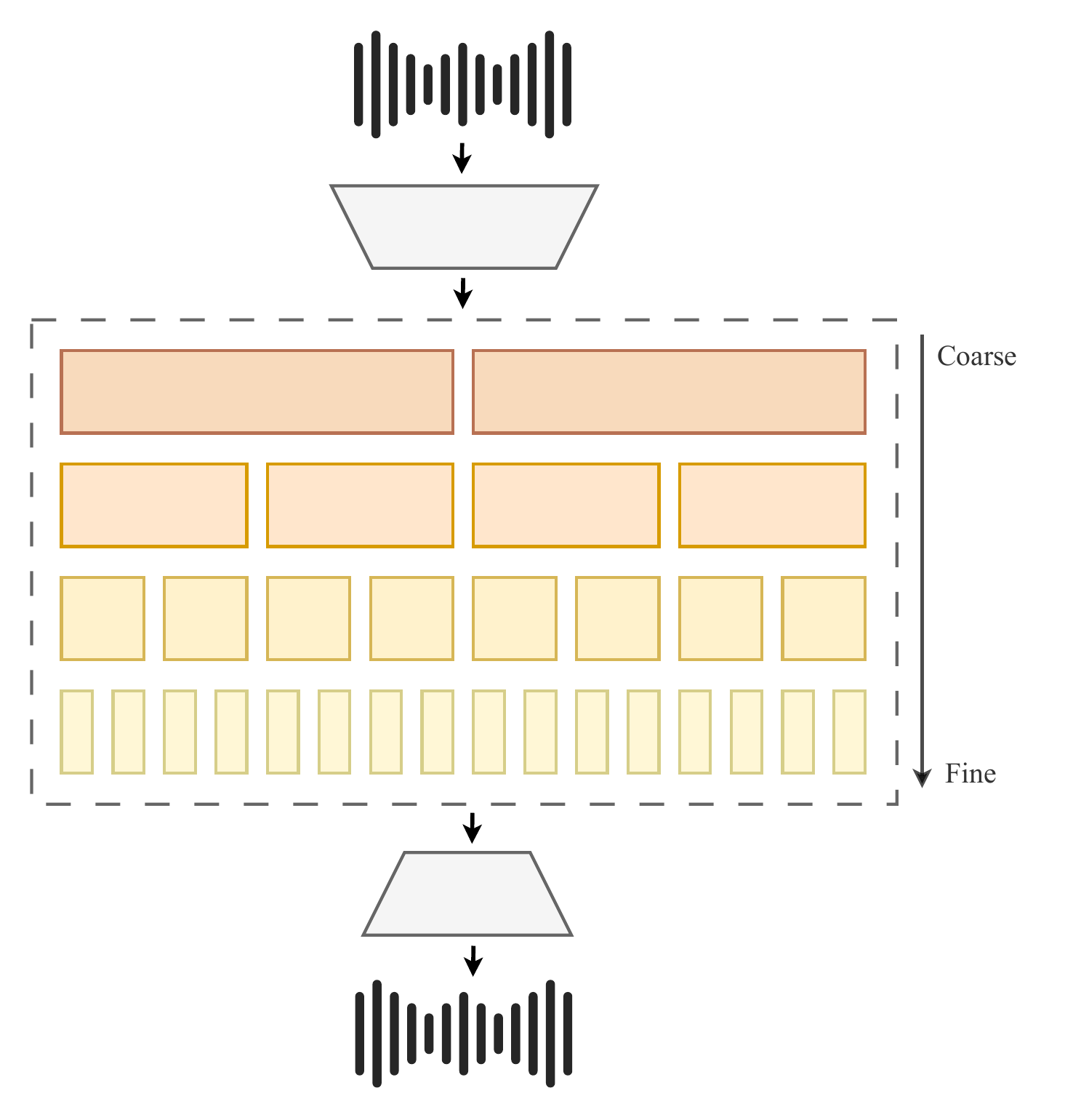}
        \caption{Multi-scale Residual Vector Quantization}
        \label{fig:rvq2}
    \end{subfigure}
    \caption{Comparison between traditional Residual Vector Quantization (RVQ) and our proposed Multi-Scale Residual Vector Quantization. The figures depict the discrete tokens produced by both methods. In the traditional RVQ approach, tokens are generated at a fixed temporal resolution, whereas SNAC utilizes a hierarchy of quantizers operating at multiple temporal resolutions, allowing the codec to capture coarse and fine details more efficiently.}
    \label{fig:snac}
\end{figure}

Our model builds upon RVQGAN \cite{kumar2024high}, an encoder-decoder network with Residual Vector Quantization (RVQ) in the bottleneck. It uses a cascade of $N_q$ vector quantization layers, where each layer maps the residual $\mathbf{x} \in \mathbb{R}^{T \times C}$ to a sequence of one-hot vectors of shape $T \times D$, where $T$ denotes the number of frames, $C$ is the encoder dimension, and $D$ is the codeword dimension.

\paragraph{Multi-Scale Residual Vector Quantization}
Our work extends RVQGAN by introducing Multi-scale Residual Vector Quantization (illustrated in Figure~\ref{fig:snac}). At each iteration $i$, we downsample the residuals by a factor of $W_i$, perform codebook lookup, and then upsample by the same factor $W_i$ to match the original temporal resolution $T$ of $\mathbf{x}$. In practice, we use average pooling for downsampling and nearest-neighbor interpolation for upsampling.

\paragraph{Noise Block}

To introduce stochasticity and enhance the decoder's expressiveness, we add a \emph{Noise Block} after each upsampling layer. This block adds noise to the activations by updating the input: 
$\mathbf{x} \leftarrow \mathbf{x} + \text{Linear}(\mathbf{x}) \odot \boldsymbol{\epsilon},$ where $\boldsymbol{\epsilon} \sim \mathcal{N}(0, 1)$ is Gaussian noise, and $\odot$ denotes element-wise multiplication. This mechanism allows the model to inject input-dependent noise. We find that the Noise Block improves reconstruction quality and leads to better utilization of the codebook.

\paragraph{Depthwise Convolution }

Depthwise separable convolutions were introduced to create lighter models for vision applications \cite{MobileNets}. By applying a single filter to each input channel, this method significantly reduces computation and model size. We propose using depthwise convolutions in the generator to not only decrease the number of parameters but also to stabilize training. GAN-based vocoders are notoriously unstable, often experiencing divergent gradients early in training which can lead to instabilities and even model collapse \cite{leebigvgan}.

\paragraph{Local Windowed Attention}

In our model, we include a single layer of local windowed attention \cite{beltagy2020longformer} at the lowest temporal resolution in both the encoder and decoder. This is motivated by the attention mechanism's ability to adaptively focus on relevant features across different inputs. Also it can complement the subsequent average pooling by helping to capture contextual representations. Similarly, \cite{defossez2022high} integrates LSTM layers to model temporal dependencies more effectively.

\section{Experiments}

We carried out two studies: one where we trained the SNAC codec for general audio -- covering music, sound effects, and environmental sounds -- and another focused on speech. The speech-specific study aimed to maintain quality with lower bitrates and reduced compute. This approach was inspired by traditional audio codecs that offer specialized algorithms for speech coding \cite{rfc6716}.

\subsection{Model architecture}
\label{sec:model_architecture}

\paragraph{General Audio}

Our model follows the encoder-decoder architecture described in \cite{kumar2024high}, with additional noise blocks after each transposed convolution in the decoder. Both the encoder and decoder incorporate local windowed attention layers at the lowest temporal resolution. We replace most convolutions with depthwise convolutions, except in the embedding, output projection, and upsample layers. The encoder uses a cascade of downsampling layers with rates of [2, 3, 8, 8], with corresponding upsampling layers in the decoder at rates [8, 8, 3, 2]. In the RVQ, we use the downsample factors (strides) of [8, 4, 2, 1], effectively compressing a 44.1 kHz input signal into four sequences of tokens at rates of 14, 29, 57, and 115 Hz. Each codebook holds 4096 entries (12-bit), leading to a total bitrate of 2.6 kbps. The model consists of 16 M parameters in the encoder and 38.3 M in the decoder, totaling 54.5~M parameters.
We apply the same architecture to train on 32 kHz audio, resulting in token rates of 10, 21, 42, and 83 Hz, with a total bitrate of 1.9 kbps.

\paragraph{Speech}

For the speech codec, we modify the architecture by adjusting the downsampling factors in the encoder (and correspondingly, the decoder) to [2, 4, 8, 8]. In the residual vector quantization, we use strides of [4, 2, 1]. This model is trained on 24 kHz audio, resulting in token rates of 12, 23, and 47 Hz, with an effective bitrate of 984 bits per second. Additionally, we reduce the number of convolution channels, resulting in 6.7 million parameters in the encoder and 13.0 million in the decoder, for a total of 19.8 million parameters. We omit local windowed attention layers in the speech codec, making the architecture purely convolutional.

\subsection{Training}

Our training largely follows the GAN setup described in \cite{kumar2024high} with a multi-period discriminator \cite{kong2020hifi} and a complex multi-scale STFT discriminator \cite{jang2021univnet}. Notably, thanks to depthwise convolutions, we are able to use a higher initial learning rate for AdamW at 6e-4, decaying at a rate of $\lambda=0.999994$ per iteration. We do not use gradient clipping, as the training remains stable.

The models are trained for 800k iterations, with each batch consisting of 16 audio clips of approximately 0.8 seconds each. This results in a total of 2730 hours of unique audio samples, sampled without replacement from our internal dataset. We train the SNAC 32 kHz and 44 kHz versions on a general audio dataset with sampling weights of 80\% music, 10\% sound effects, and 10\% speech. The SNAC 24 kHz version is trained exclusively on a speech dataset. However, it is important to note that this dataset is diverse and may still expose the model to non-speech sounds, such as podcast intro music, jingles, or various background noises and effects.

\subsection{Ablation Study}

To evaluate the impact of the proposed components, we conducted an ablation study on a 44 kHz version of our model, trained on a general audio dataset. We used a smaller batch size of 8 samples and trained each variant for 250k iterations on the same subset of audio files. The results, summarized in Table~\ref{tab:ablation_study}, demonstrate the effectiveness of each component.

The \emph{Single-scale} variant represents a typical RVQ setup, where a single temporal resolution is used for all VQ codebooks. All other modules, including noise blocks, local attention, depthwise convolutions, and training hyperparameters, remain consistent with the baseline. In this setup, the encoder downsample rates are set to [2, 4, 8, 8], and the model uses 3 levels of RVQ, each with 4096 codebook entries, resulting in a total bitrate of 3.1 kbps. Despite the larger bitrate compared to the SNAC baseline (2.6 kbps), this variant performs worse across all objective metrics, highlighting the importance of our multi-scale quantization approach for audio.

The Noise Block leads to a notable improvement in reconstruction quality, especially in terms of the SI-SDR score. In contrast, the local attention layers provide only marginal gains. We hypothesize that deeper attention networks with larger context windows might yield more meaningful contextual representations. However, these also increase computational costs and latency. For our lightweight speech codec, we chose to omit the local attention layers. Additionally, when training with standard (non-depthwise) convolutions using the same hyperparameters as the baseline, we observed unstable training and eventual model collapse.

\begin{table}[t]
\centering
\caption{Ablation study results comparing various model variants. \emph{Single-scale RVQ} uses a single temporal resolution for all VQ codebooks. \emph{w/o Noise Block} is the model without the noise blocks in the decoder. \emph{w/o LWA} indicates the model without local attention layers. \emph{w/o DW Conv.} refers to the model without depthwise convolutions, which failed to produce stable results (metrics not available). The results highlight the effectiveness of our multi-scale quantization approach and the contributions of each component to overall performance.}
\label{tab:ablation_study}
\begin{tabular}{@{}lcccc@{}}
\toprule
\textbf{Model Variant} & \textbf{ViSQOL (↑)} & \textbf{SI-SDR (↑)} & \textbf{Mel Distance (↓)} & \textbf{STFT Distance (↓)} \\ 
\midrule
SNAC (baseline) & \textbf{4.00} & \textbf{3.95} & \textbf{1.38} & \textbf{1.62} \\
Single-scale RVQ & 3.89 & 3.73 & 1.44 & 1.66 \\
w/o Noise Block & 3.94 & 3.32 & 1.43 & 1.65 \\
w/o LWA & 3.99 & 3.90 & 1.39 & 1.63 \\
w/o DW Conv. & N/A & N/A & N/A & N/A \\
\bottomrule
\end{tabular}
\end{table}

\subsection{Evaluation}

We evaluate SNAC using a combination of objective metrics and a MUSHRA-like (Multiple Stimuli with Hidden Reference and Anchor)~\cite{series2014method} listening study to assess the perceived quality of speech and music samples generated by different codecs. For the speech evaluation, we used the DAPS dataset~\cite{daps}, selecting 10 samples of varying lengths (5–9 seconds), with five female voices and five male voices.
For the music evaluation, we used 10 samples trimmed to 5 seconds from the MUSDB18-HQ dataset~\cite{musdb18-hq}. In addition to the listening study, we evaluated the codecs on commonly used objective metrics to quantitatively assess signal quality and speech intelligibility.

\begin{figure}
  \centering
  \begin{subfigure}{0.5\textwidth}
    \centering
    \includegraphics[height=6.8cm]{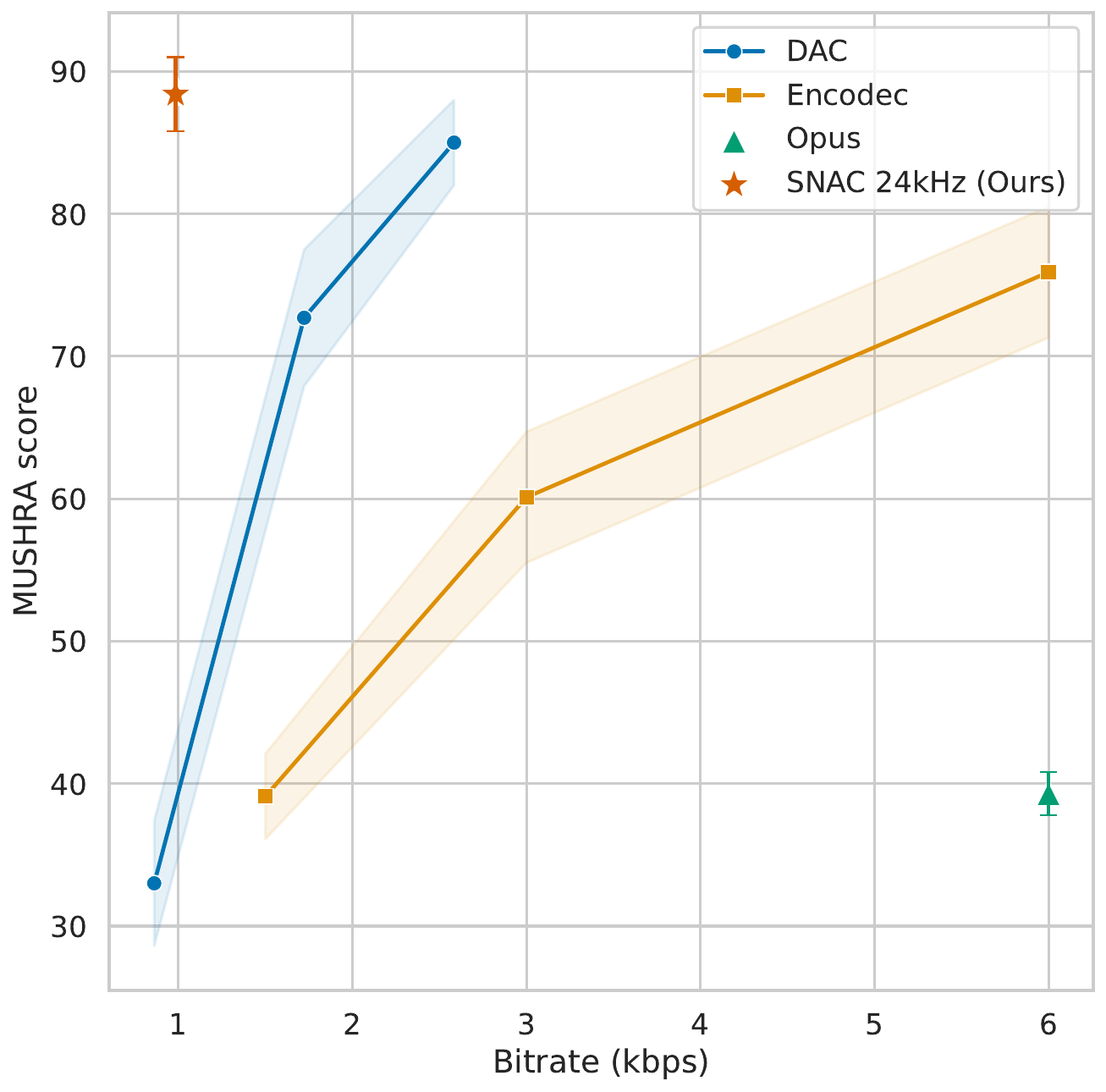}
    \caption{Speech evaluation}
    \label{fig:sub1}
  \end{subfigure}
  \hfill
  \begin{subfigure}{0.49\textwidth}
    \centering
    \includegraphics[height=6.8cm]{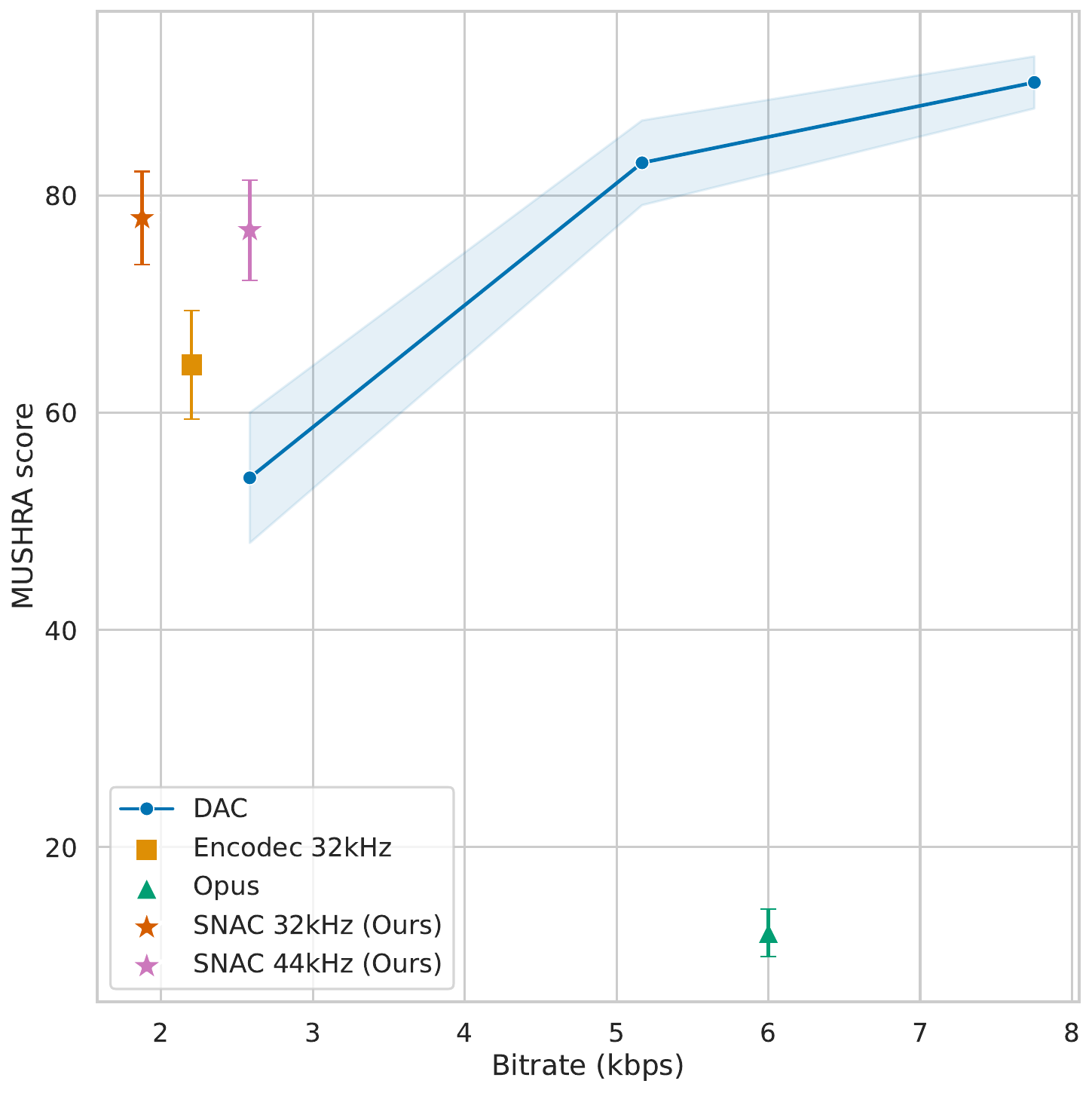}
    \caption{Music evaluation}
    \label{fig:sub2}
  \end{subfigure}
  \caption{Results of the MUSHRA listening study with 95\% confidence intervals. We visualize the performance of SNAC compared to previous state-of-the-art approaches. We find that SNAC outperforms existing speech codecs while using a significantly lower bitrate and performs comparably to DAC in music reconstruction quality at a considerably lower bitrate.}
  \label{fig:human_eval}
\end{figure}

Participants in the MUSHRA test were presented with a hidden reference (the original, unprocessed audio), an anchor (compressed using the Opus codec at 6 kbps, with the corresponding audio and speech preset), and samples that had been encoded and decoded by various neural audio codecs.
All samples were loudness normalized to -12dB LUFS.
Ratings were collected using the webMUSHRA framework~\cite{schoeffler2018webmushra} from a group of audio experts.
Results of the human evaluation are presented in Figure \ref{fig:human_eval}, and the full results, including objective metrics, are presented in Table~\ref{tab:speech_table} and Table~\ref{tab:music_table}.

\paragraph{Music}
We compare the two SNAC variants for general audio introduced in Section~\ref{sec:model_architecture} with the 32 kHz checkpoint of Encodec~\cite{defossez2022high} from MusicGen \cite{musicgen} and the official DAC \cite{kumar2024high} checkpoint using 3, 6, or 9 codebooks.  
We observe that SNAC significantly outperforms other codecs, such as Encodec (32 kHz) and DAC (with 3 codebooks), which operate at comparable bitrates. Notably, SNAC even competes with codecs operating at more than twice its bitrate. Furthermore, the difference in perceived audio quality between the SNAC models at 32 kHz and 44 kHz is marginal, suggesting that the 32 kHz model is adequate for most tasks, offering the added benefit of a lower bitrate.

\paragraph{Speech}
For speech, we compare the SNAC speech model with EnCodec (24 kHz checkpoint) and DAC using varying numbers of codebooks. In our evaluation, SNAC consistently outperforms all other codecs. Notably, even at bitrates below 1 kbit/s, SNAC maintains audio quality that closely approaches the reference signal. This efficiency makes it particularly advantageous for bandwidth-constrained applications, where preserving intelligibility and clarity of voices is crucial. 

\section{Conclusion}

We introduced the Multi-Scale Neural Audio Codec (SNAC), an extension of Residual Vector Quantization that uses quantizers operating at multiple temporal resolutions. This multi-scale approach adapts to the inherent structure of audio signals, leading to more efficient compression. Ablation studies confirmed the significance of our design choices. SNAC outperformed existing state-of-the-art codecs in both music and speech domains, delivering higher audio quality at lower bitrates, as demonstrated by extensive objective and subjective evaluation. By open-sourcing our code and models, we aim to contribute to the advancement of neural audio compression research.

\clearpage
\bibliography{neurips_2024}
\bibliographystyle{unsrtnat}

\clearpage
\appendix
\section{Full results}

\begin{table}[h]
\centering
\caption{Comparison of SNAC with previous neural audio codecs on speech samples. The bitrate is measured in kbps, Mel and STFT measure the L1 distance between the reconstruction and the ground-truth. MUSHRA scores are reported with a 95\% confidence interval, and the reference signal in MUSHRA was rated at $99.5\pm0.3$.}
\label{tab:speech_table}
\begin{tabular}{@{}llccccc@{}}
\toprule
\textbf{Model} & \textbf{Bitrate} & \textbf{ViSQOL (↑)} & \textbf{SI-SDR (↑)} & \textbf{Mel (↓)} & \textbf{STFT (↓)} & \textbf{MUSHRA (↑)} \\
\midrule
Opus & 6 & 3.84 & 2.28 & 4.99 & 4.01 & $39.3\pm1.5$ \\
\midrule
EnCodec & 6 & 4.36 & 6.83 & 1.60 & 1.81 & $75.9\pm4.6$ \\
 & 3 & 4.12 & 3.72 & 1.80 & 1.94 & $60.1\pm4.6$ \\
 & 1.5 & 3.75 & 0.85 & 2.05 & 2.07 & $39.1\pm3.0$ \\
\midrule
DAC & 2.5 & 4.28 & 6.43 & 1.37 & 1.65 & $85.0\pm3.0$ \\
 & 1.7 & 4.03 & 4.22 & 1.56 & 1.74 & $72.7\pm4.8$ \\
 & 0.8 & 3.49 & -0.16 & 2.03 & 1.96 & $33.0\pm4.4$ \\
\midrule
SNAC (ours) & 0.98 & 4.14 & 0.82 & 1.50 & 1.78 & $88.4\pm2.6$ \\
\bottomrule
\end{tabular}
\end{table}

\begin{table}[h]
\centering
\caption{Comparison of SNAC with previous neural audio codecs on music samples. The bitrate is measured in kbps, Mel and STFT measure the L1 distance between the reconstruction and the ground-truth. MUSHRA scores are reported with a 95\% confidence interval, and the reference signal in MUSHRA was rated at $99.8\pm0.3$.}
\label{tab:music_table}
\begin{tabular}{@{}llccccc@{}}
\toprule
\textbf{Model} & \textbf{Bitrate} & \textbf{ViSQOL (↑)} & \textbf{SI-SDR (↑)} & \textbf{Mel (↓)} & \textbf{STFT (↓)} & \textbf{MUSHRA (↑)} \\
\midrule
Opus & 6 & 1.54 & 2.24 & 7.33 & 5.71 & $12.1\pm2.2$ \\
\midrule
EnCodec & 2.2 & 3.66 & 5.41 & 1.91 & 1.97 & $64.4\pm5.0$ \\
\midrule
DAC & 7.74 & 4.19 & 10.60 & 1.15 & 1.50 & $90.4\pm2.4$ \\
 & 5.16 & 4.09 & 8.31 & 1.30 & 1.56 & $83.0\pm3.9$ \\
 & 2.5 & 3.91 & 5.04 & 1.54 & 1.69 & $54.0\pm6.0$ \\
\midrule
SNAC (ours) & 1.9 & 3.79 & 4.01 & 1.75 & 1.89 & $77.9\pm4.3$ \\
 & 2.6 & 4.04 & 5.17 & 1.42 & 1.59 & $76.8\pm4.6$ \\
\bottomrule
\end{tabular}
\end{table}

\end{document}